\documentclass[prd,onecolumn,notitlepage,amsmath,amssymb,floatfix,superscriptaddress]{revtex4-1}
\usepackage{graphicx}
\usepackage{amssymb}
\usepackage{amsmath}
\usepackage{hyperref}
\usepackage{bm}
\usepackage{color}
\DeclareMathAlphabet\mathbfcal{OMS}{cmsy}{b}{n}

\definecolor{darkred}{cmyk}{0,1.00,1.00,0.3}
\definecolor{purple}{cmyk}{0.5,1.0,0,0} 
\definecolor{darkgreen}{cmyk}{0.85,0.2,1.00,0.2} 

\newcommand{\mnras}{Mon.~Not.~Roy.~Astron.~Soc.}
\newcommand{\physrep}{Phys.~Rep.}
\newcommand{\jcap}{JCAP}
\newcommand{\pasj}{PASJ}

\newcommand{\apjs}{\apj~Suppl.}
\newcommand{\aap}{Astronomy \& Astrophysics}

\newcommand{\bbr}{{\bf r}}
\newcommand{\bx}{{\bf x}}
\newcommand{\by}{{\bf y}}
\newcommand{\bk}{{\bf k}}

\newcommand{\bq}{{\bf q}}

\newcommand{\hk}{\hat{k}}

\newcommand{\lin}{{\rm L}}

\newcommand{\tdelta}{\tilde{\delta}}
\newcommand{\tW}{\tilde{W}}

\newcommand{\br}{{\rm b}}

\renewcommand{\O}{\mathcal{O}} 

\def\avrg#1{\left\langle #1 \right\rangle}

\begin{document}

\title[Super-Survey Tidal Effect]{Large-scale tidal effect on
redshift-space power spectrum in a finite-volume survey}

\author{Kazuyuki Akitsu}
\affiliation{Kavli Institute for the Physics and Mathematics of the Universe
(WPI), The University of Tokyo Institutes for Advanced Study (UTIAS),
The University of Tokyo, Chiba 277-8583, Japan}
\affiliation{Department of Physics, Graduate School of Science, The University of
Tokyo, 7-3-1 Hongo, Bunkyo-ku, Tokyo 113-0033, Japan}
\author{Masahiro Takada}
\affiliation{Kavli Institute for the Physics and Mathematics of the Universe
(WPI), The University of Tokyo Institutes for Advanced Study (UTIAS),
The University of Tokyo, Chiba 277-8583, Japan}
\author{Yin Li}
\affiliation{Kavli Institute for the Physics and Mathematics of the Universe
(WPI), The University of Tokyo Institutes for Advanced Study (UTIAS),
The University of Tokyo, Chiba 277-8583, Japan}
\affiliation{Berkeley Center for Cosmological Physics, Department of Physics,
and Lawrence Berkeley National Laboratory, University of California, Berkeley, CA 94720, USA}

\begin{abstract}
Long-wavelength matter inhomogeneities contain cleaner information on
the nature of primordial perturbations as well as the physics of the
early universe.  The large-scale coherent overdensity and tidal force,
not directly observable for a finite-volume galaxy survey, are both
related to the Hessian of large-scale gravitational potential and
therefore of equal importance. We show that the coherent tidal force
causes a homogeneous anisotropic distortion of the observed distribution
 of galaxies in all three directions,  perpendicular and parallel  to the
line-of-sight direction.  This effect mimics the redshift-space
distortion signal of galaxy peculiar velocities, as well as a distortion
by the Alcock-Paczynski effect.  We quantify its impact on the
redshift-space power spectrum to the leading order, and discuss its
importance for the ongoing and upcoming galaxy surveys.
\end{abstract}

\maketitle

\section{Introduction}

Observations of large-scale structure in the universe through a
wide-area spectroscopic survey of galaxies are a very powerful probe of
fundamental physics, e.g. to test the nature of dark energy via the
baryon acoustic oscillation (BAO) measurements of cosmological distances
\citep{SeoEisenstein:03,HuHaiman:03,Eisensteinetal:05,Aubourgetal:15},
to test the gravity theory on cosmological scales \citep{Zhangetal:07},
to weigh the neutrino mass \citep{Huetal:98,Takadaetal:06,Saitoetal:08},
to extract the physics of the early universe
\cite{Dalaletal:08,Carboneetal:11,ArkaniHamedMaldacena:15}, to constrain
the spatial curvature \citep{TakadaDore:15}, and to constrain the abundance of
light relics such as axions \citep{Hlozeketal:14}. The current-generation
galaxy surveys such as the SDSS Baryon Oscillation Spectroscopic Survey
(BOSS) have provided stringent cosmological constraints that are yet
complementary to constraints from the cosmic microwave background (CMB)
\citep{Reidetal:12,Alametal:16}. There are upcoming wide-area galaxy
surveys probing the three-dimensional distribution of galaxies at higher redshifts:
the Subaru Prime Focus Spectrograph (PFS)
\citep{Takadaetal:14}, the Dark Energy Spectrograph Instrument (DESI)
\citep{DESI:16}, the ESA Euclid
\footnote{\url{http://sci.esa.int/euclid/}}, the NASA SPHEREx
\citep{SPHEREX:14} and the NASA WFIRST-AFTA \citep{WFIRST:15}.

To attain the full potential of wide-area galaxy surveys, it is crucial
to understand the statistical properties of large-scale structure
probes. Even though the initial density field is nearly Gaussian, the
subsequent nonlinear evolution of structure formation causes substantial
non-Gaussian features in the observed distribution of galaxies and
matter \citep{Bernardeauetal:02}.  Most of the useful cosmological
information lies in the weakly or deeply nonlinear regime,
where different Fourier modes are no longer independent but tightly coupled.

The fact that any galaxy survey has to be done within a finite volume
also causes an unavoidable uncertainty in the actual cosmological
analysis.  Matter density perturbations with very long wavelengths
outside a survey volume, hereafter called super-survey modes, should be
present, but are not directly observable.
In the nonlinear regime of structure formation, 
the super-survey modes become coupled to short-wavelength modes
inside the survey volume.
Consequently
cosmological probes measured from a given survey region are modulated
coherently by the super-survey modes, and
the effects need to be taken into account in the analysis in order not
to have any bias in cosmological parameter estimation. In addition the
super-survey modes are tricky to consider, because their effects vanish
for $N$-body simulations with periodic boundary conditions that have no
contribution of modes outside the simulation box.

Various works have studied the super-survey effects for cosmological
observables such as the weak lensing correlation functions
\citep{HuKravtsov:03,Hamiltonetal:06,TakadaBridle:07,TakadaJain:09,Satoetal:09,SherwinZaldarriaga:12,Kayoetal:13,TakadaHu:13,TakadaSpergel:13,Schaanetal:14,Lietal:14a,Lietal:14b,MohammedSeljak:14,Mohammedetal:16,Daietal:15,Shirasakietal:16}.
Most of them focused on the effects of the large-scale coherent
overdensity, denoted by $\delta_{\br}$ \citep[see][for a unified
formulation of the effect]{TakadaHu:13}.  The effect of $\delta_\br$ on
sub-survey modes for a cold dark matter model with the cosmological
constant ($\Lambda$CDM) can be absorbed into an apparent curvature
parameter of the local volume -- a separate universe picture
\citep{McDonald03,Sirko:05,Gnedinetal:11,Baldaufetal:11,Lietal:14a,Wagneretal:15,Daietal:15}.
This approach allows one to include the fully nonlinear mode-coupling of
$\delta_\br$ with all short-wavelength modes, by performing $N$-body
simulations on a perturbed background correctly capturing the local
expansion.

However, the effects of a long-wavelength and coherent gravitational tidal
force on short-wavelength modes have yet to be fully studied. The
coherent overdensity and the coherent tidal force are both related to
the Hessian of the gravitational potential
(or more generally the metric perturbations), and have comparable
amplitudes in each realization.  Since the long-wavelength tidal field
could have a direct link to the physics of the early universe
\citep[e.g.][]{Erickceketal:08,Whiteetal:13,Creminellietal:13,ArkaniHamedMaldacena:15},
it would be interesting to explore the effects from the observed galaxy
distribution.  Recently \citet{IpSchmidt:16} developed a formulation to
describe effects of the coherent tidal force on nonlinear structure
formation in a local volume within the framework of general relativity
\citep[also see][]{EisensteinLoeb:95,HuiBertschinger:96,BondMyers:96}.
In this paper we study how the super-survey coherent tidal force causes
an apparent anisotropic clustering in the galaxy distribution.  We will
show that the effects appear to look like the redshift-space distortion
due to peculiar motions of galaxies as well as the Alcock-Paczynski
effect.

The structure of this paper is as follows. In \S~\ref{sec:formulation}
we derive a formula to describe an effect of the large-scale coherent
gravitational tidal force on the redshift-space galaxy power spectrum
measured in a given realization of a finite-volume survey,
followed by its contribution to the covariance matrix of the
quadrupole power spectrum.
In \S~\ref{sec:results}, we assess its impact on the quadrupole power
spectrum measurement for a hypothetical galaxy survey.
\S~\ref{sec:discussion} is devoted to discussion. In
Appendix~\ref{app:ssc} we derive the response of the power spectrum to
the large-scale tide, based on the perturbation theory.

\section{Super-survey tidal effect}
\label{sec:formulation}

\subsection{Super-survey modes}
\label{sub:modes}

For purpose of the following discussion let us consider the
gravitational potential field smoothed with a survey window function:
\begin{equation}
 \Phi_W(\bx)\equiv \frac{1}{V_W}\int\!\mathrm{d}^3\by~ \Phi(\by)W(\by-\bx),
\label{eq:PhiW_def}
\end{equation}
where $V_W=\int\!\mathrm{d}^3\by~W(\by-\bx)$.
For simplicity throughout the paper we assume a connected survey geometry,
which does not have any hole or masked region.
The survey window thus defines the boundary of a survey
region around the fiducial point $\bx$; $W(\by-\bx)=1$ if the vector
$\by-\bx$ is inside a survey region, otherwise $W(\by-\bx)=0$.  In this
way we can consider $\Phi_W(\bx)$ as the smoothed gravitational field
as a function of the position $\bx$.
If a typical length scale
of the survey window is $L$, the above integration smooths out all
fluctuations with scales smaller than $L$ around the position $\bx$.
$\Phi_W(\bx)$ only varies significantly on scales comparable to or
greater than $L$.

Now suppose that a hypothetical survey region is located at the position
$\bx_0$.  Then consider to Taylor-expand the smoothed gravitational
field around the position $\bx_0$ as
\begin{eqnarray}
\Phi_W(\bx)&=&\Phi_W(\bx_0)+\left.\nabla_i \Phi_W\right|_{\bx_0}x^i
+\frac{1}{2}\left.\nabla_i\nabla_j \Phi_W\right|_{\bx_0}x^ix^j +
\O(\nabla^3\Phi_W|_{\bx_0} x^3)\nonumber\\
&=&\Phi_W(\bx_0)+\left.\nabla_i \Phi_W\right|_{\bx_0}x^i
 +\frac{2}{3}\pi G\bar{\rho}_{\rm m}a^2\left.\delta_{\rm b}\right|_{\bx_0}x^2
 +2\pi G\bar{\rho}_{\rm m}a^2
 \left.\tau_{Wij}\right|_{\bx_0}x^ix^j+
 \O(\nabla^3\Phi_{W}|_{\bx_0}x^3),
 \label{eq:Phi_exp}
\end{eqnarray}
%
where the comoving displacement $x^i\equiv (\bx-\bx_0)^i$,
$\nabla_i\equiv\partial/\partial x^i$,
$a(t)$ is the scale factor of the global universe,
and $\delta_{\br}$ is the smoothed overdensity in the survey window (see below).
We have used the Poisson equation, $\Delta \Phi(\bx)=4\pi
G\bar{\rho}_{\rm m}a^2\delta(\bx)$.  $\tau_{Wij}$ is the
smoothed tidal field defined as the traceless Hessian matrix
of the smoothed gravitational field
\begin{equation}
 \tau_{Wij}\equiv \frac{1}{4\pi G\bar{\rho}_{\rm m}a^2}\left(
  \Phi_{W,ij}-\frac{1}{3}\delta^K_{ij}\Delta\Phi_W\right),
\label{eq:tau_def}
\end{equation}
and $\delta^K_{ij}$ is the Kronecker delta function.
We introduced the prefactor
$(1/4\pi G\bar{\rho}_{\rm m}a^2)$ in the definition of $\tau_{Wij}$ to make it
dimensionless.  By using the properties of survey window
as well as the {partial integral,}
we can rewrite the partial derivatives of the smoothed gravitational field,
for example, as 
\begin{eqnarray}
 \left.\nabla_i\Phi_W\right|_{\bx_0}&\equiv &\frac{\partial }{\partial  x^i}
  \left[\frac{1}{V_W}
   \int\!\mathrm{d}^3\by~\Phi(\by)W(\by-\bx)
      \right]_{\bx_0}
  =
  \left.\frac{1}{V_W}
\int\!\mathrm{d}^3\by~\Phi(\by)\frac{\partial W(\by-\bx)}{\partial x^i}
				\right|_{\bx_0}\nonumber\\
 &=& 
    \left.\frac{1}{V_W}
\int\!\mathrm{d}^3\by~\Phi(\by)(-1)\frac{\partial W(\by-\bx)}{\partial y^i}
							  \right|_{\bx_0}\nonumber\\
 &=& 
   { \frac{1}{V_W} \left[  \int\!\mathrm{d}^3\by~ \frac{\partial }{\partial y^i} \left\{ \Phi(\by)(-1)W(\by-\bx) \right\} - \int\!\mathrm{d}^3\by~  \frac{\partial \Phi(\by) }{\partial y^i}(-1)W(\by-\bx)  \right]_{\bx_0} } \nonumber \\ 
 &=& \left.\frac{1}{V_W}
\int\!\mathrm{d}^3\by~\frac{\partial \Phi(\by)}{\partial y^i} W(\by-\bx)
							  \right|_{\bx_0}=\left.\Phi_{W,i}\right.|_{\bx_0}. 
\end{eqnarray}
%
That is, the derivatives of the smoothed field in Eq.~(\ref{eq:Phi_exp})
are equivalent to the survey window average of the derivatives of the
gravitational potential field. With this equality, we rewrote the
Laplacian of the smoothed field in the third line on the right hand side of
Eq.~(\ref{eq:Phi_exp}) as
\begin{eqnarray}
 \left.\Delta\Phi_W\right|_{\bx_0}&=&\left.\frac{1}{V_W}\int\!\mathrm{d}^3\by~
  \Delta\Phi(\by)W(|\by-\bx|)\right|_{\bx_0}\nonumber\\
 &=&\left.\frac{1}{V_W}\int\!\mathrm{d}^3\by~4\pi G \bar{\rho}_{\rm m}a^2 \delta(\by)
  W(|\by-\bx|)\right|_{\bx_0}=4\pi G\bar{\rho}_{\rm m}a^2\delta_{\rm b}|_{\bx_0}.
\end{eqnarray}
where $\delta_\br(\bx_0)\equiv (1/V_W)\int\!\!\mathrm{d}^3\bx~W(\by-\bx_0)\delta(\by)$.
All the coefficients of $x^n$ on the right hand side of Eq.~(\ref{eq:Phi_exp})
are evaluated at the position $\bx_0$, at a given time,
$\delta_\br=\delta_\br(t)$ and $\tau_{Wij}=\tau_{Wij}(t)$.
Hereafter we will often omit the dependence of $\bx_0$ in the
super-survey modes when considering a fixed position of the survey
region. 
As long as the survey window is sufficiently large, the super-survey
modes evolve linearly, i.e. $\delta_{\br}\propto
D(t)$ and $\tau_{Wij}\propto \Phi_W/(\bar{\rho}_{\rm m}a^2)\propto D(t)$,
where $D(t)$ is the linear growth function \citep{DodelsonBook}.

The ensemble averages of the super-survey modes,
which are equivalent to the average when varying the position $\bx_0$
for a fixed survey window function, can be estimated based on the linear
theory for an assumed $\Lambda$CDM model.
For a general survey window
$\avrg{\Phi_W}=\avrg{\tau_{Wij}}=\avrg{\delta_{\br}}=0$ and their
variances are expressed as
\begin{eqnarray}
&& \sigma_\br^2\equiv  \avrg{\delta_\br^2}=\frac{1}{V_W^2}\int\!\!\frac{\mathrm{d}^3\bq}{(2\pi)^3}~P^{L}(q)|\tW(\bq)|^2,
 \nonumber\\
&& (\sigma_{\tau_{ij}\tau_{lm}})^2\equiv  \avrg{\tau_{Wij}\tau_{Wlm}}
 =\frac{1}{V_W^2}\int\!\!\frac{\mathrm{d}^3\bq}{(2\pi)^3}~
 \left(\hat{q}_i\hat{q}_j-\frac{\delta^K_{ij}}{3}\right)
\left(\hat{q}_l\hat{q}_m-\frac{\delta^K_{lm}}{3}\right)
  P^L(q)|\tW(\bq)|^2,
\label{eq:sigma_tau}
\end{eqnarray}
where $\hat{q}_i\equiv q_i/q$, we have used
$\avrg{\tdelta_{\bk}\tdelta_{\bk'}}\equiv(2\pi)^3P^L(k)\delta_D^3(\bk+\bk')$
as well as the Poisson equation in the Fourier space,
$-k^2\tilde{\Phi}_\bk=(4\pi G\bar{\rho}_{\rm m}a^2)\tdelta_\bk$, and
$P^L(k)$ is the linear mass power spectrum. For a general survey window,
$\avrg{\delta_\br\tau_{Wij}}\neq 0$.  The linear variance $\sigma_\br$
and $\avrg{\tau_{Wij}\tau_{Wlm}}$ can be easily computed for any survey
geometry, either by evaluating Eq.~(\ref{eq:sigma_tau}) directly or
using Gaussian realizations of the linear density field. Note that, even
for a fixed survey volume, different components of the linear tidal
variance $\avrg{\tau_{Wij}\tau_{Wlm}}$ generally have different
amplitudes for an irregular-shaped window; for example, if a survey
window has a collapsed shape rather than an isotropic shape,
$\avrg{\tau_{Wij}\tau_{Wlm}}$ has a greater amplitude for the components
corresponding to the smaller window size (see below for further
discussion).

For an isotropic window, $\tW(\bq)=\tW(q)$, the components of the linear
tidal variance are simplified as
\begin{eqnarray}
  &&\avrg{\delta_\br\tau_{Wij}}=0, \nonumber\\
  &&\sigma_{\tau}^2\equiv \avrg{(\tau_{W11})^2}=
  \avrg{(\tau_{W22})^2}=\avrg{(\tau_{W33})^2}=\frac{3}{4}\avrg{(\tau_{Wij})^2}_{i\ne
  j}=
  \frac{4}{45V_W^2}\int\!\!\frac{q^2\mathrm{d}q}{2\pi^2}~P^{L}_\delta(q)|\tW(q)|^2
  =\frac{4}{45}\sigma_\br^2,
  \label{eq:sigma_tau_iso}
\end{eqnarray}
and other variances are vanishing: $\avrg{\tau_{Wij}\tau_{Wlm}}=0$.
Thus the large-scale overdensity and tidal variances have comparable
amplitudes, $\sigma_\tau\sim \sigma_\br/3$, because both are related to 
the Hessian of the gravitational field.
No correlation between $\delta_\br$
and $\tau_{Wij}$ means that the two carry independent information of the
super-survey modes.
The higher-order derivatives than $\Phi_{W,ijl}$ in
Eq.~(\ref{eq:Phi_exp}) are more sensitive to smaller scale modes
(larger-$k$ modes), and are suppressed by a factor of $(x/L)^n$, where
$x$ is the length scale of sub-survey modes we are interested in and $L$
is the survey size. Hence $\delta_\br$ and $\tau_{Wij}$ give
leading-order contributions to the super-survey effects.  {In fact
\citet{Lietal:14a} showed that the higher-order contributions with $n\ge
3$ seem negligible for the matter power spectrum, using the separate
universe simulations.}

\subsection{The redshift-space power spectrum}
\label{sub:Pl}

Let us consider structure formation in a finite-volume survey window in
the universe. To do this we employ a ``separate universe picture''
\citep{Baldaufetal:11,Lietal:14a,Wagneretal:15,Daietal:15} -- we
consider time-evolution of motions of particles comoving with this
finite volume region in a Lagrangian picture, which is separated from
the global universe. As can be found from Eq.~(\ref{eq:Phi_exp}),
the same coherent
force arising from the large-scale gravitational field,
$\nabla_{\bx}\Phi_W(\bx)$, acts on all
particles inside the survey region.
The first term of Eq.~(\ref{eq:Phi_exp}),
$\Phi_W|_{\bx_0}$, is vanishing, so irrelevant for
$\nabla_\bx\Phi_W(\bx)$. The force from the 2nd term,
$\Phi_{W,i}|_{\bx_0}$, causes a parallel translation of all the
particles by the same amount, and does not cause any additional
clustering inside the survey region. The force arising from the 3rd and
4th terms ($\delta_\br$ and $\tau_{Wij}$) causes the leading-order
effect on which we focus in this paper. If we consider particles
that were initially co-moving with the global comoving coordinates at a
sufficiently high redshift (where $|\delta_\br|, |\tau_{Wij}|\ll 1$),
their subsequent trajectories deviate from the global comoving
coordinates as time goes by, due to the large-scale gravitational
force. That is, their equation of motion in this ``separate'' survey
region is given as
\begin{equation}
 \ddot{X}^i=-\frac{4}{3}\pi G\bar{\rho}_{\rm m}(1+\delta_{\br})X^i
  +\frac{\Lambda}{3}X^i-4\pi G\bar{\rho}_{\rm m}\tau_{Wij}X^j,
\label{eq:modFRW}
\end{equation}
where $X^i$ is the displacement vector between the two particles (the
initially co-moving particles) in the {\em physical} coordinates, and we
have taken into account the gravitational force for a background
universe, including the effect of the cosmological constant
\citep{DodelsonBook}.  The above equation (\ref{eq:modFRW}) can be
realized as a modified Friedmann-Robertson-Walker (FRW) equation that
describes an effective expansion of the local survey region due to the
presence of super-survey modes.  The term involving $\delta_{\br}$
causes a greater or smaller gravitational force relative to the FRW
background, if the survey region is embedded into a coherent over- or
under-density region ($\delta_\br>0$ or $<0$), respectively. This effect
can be absorbed by a redefinition of the background density,
$\bar{\rho}_{W{\rm m}}=\bar{\rho}_{\rm m}(1+\delta_\br)$, as can be
found from the above equation. The effect on the growth of sub-survey
modes, through the nonlinear mode coupling, can be described by
introducing an apparent curvature parameter of the order of $\delta_\br$
in the effective FRW equation of the local universe, in the separate
universe picture \citep{Lietal:14a}\citep[see
also][]{Sirko:05,Lietal:14b,Wagneretal:15}.  The term involving the
super-survey tidal tensor $\tau_{Wij}$ is a novel effect,
and causes a {\em homogeneous} anisotropic expansion due
to its tensor nature. We meant by ``homogeneous'' here that the
expansion rate between two points inside the local volume is the same or
{\em homogeneous} independently of where the two points are placed
inside the volume, as long as the two points are taken along the same
direction, as in the Hubble law.
{This {\em homogeneity} is guaranteed by the assumption that here 
we considered only up to the second order of the Taylor expansion of the gravitational potential, 
which is the leading order effects of the super-survey modes as we discussed above.}

Using the Zel'dovich
approximation \citep{Zel'dovich:70} or the linearized Lagrangian
perturbation theory \citep[e.g.][]{Matsubara:08}, the effect of
super-survey modes on the local expansion can be described by a temporal
perturbation of the comoving coordinates of the local survey region as
\begin{equation}
q_{Wi}=q_i+\Psi_{Wij}(t)q_j,
\label{eq:qW}
\end{equation}
where 
\begin{equation}
 \Psi_{Wij}(t)=\frac{\delta^K_{ij}}{3}\delta_{\br}(t) 
  +{\tau}_{Wij}(t).
\end{equation}
%
Here $q_{Wi}$ {are} the perturbed comoving coordinates in the local survey
region, and $q_i$ is the comoving coordinate of the global
background. In the following quantities with or without subscript
``$W$'' denote the quantities in the local survey volume or the global
background, respectively.  For a sufficiently high redshift,
$|\Psi_{Wij}|\ll 1$, $q_{Wi}\simeq q_i$. Hence, the Lagrangian
coordinates in the local volume can be defined by the global comoving
coordinates at sufficiently high redshift. These effects can be also
described by a modification of the scale factor of the local
background. Note that, in the separate universe picture, the physical
length scale should be kept the same in the local volume and the global
background, as discussed in \citet{Lietal:14a}:
\begin{equation}
a_W\lambda_W=a\lambda, 
\end{equation}
where $\lambda_W$ and $\lambda$ are in the comoving wavelength
scales. Hence, the effect of $\delta_\br$ can also be realized as a
modification of the scale factor: $a_W(t)\simeq
a(t)\left[1-\delta_\br(t)/3\right]$ up to the linear order of
$\delta_\br$, which reproduces the results around Eq.~(35) in
\citet{Lietal:14a}. On the other hand, the coherent tidal force
$\tau_{Wij}$ causes a homogeneous anisotropic expansion effect on the
local comoving coordinates. If we take the axes of local comoving
coordinates along the principal axes of the coherent tidal force, which
can be done without loss of generality, the tensor $\tau_{Wij}$ becomes
diagonal: $\tau_{Wij}=\tau_{Wi}\delta^K_{ij}$. Then the deformation of
the comoving coordinates can be realized as a homogeneous anisotropic
deformation of the scale factor along each axis up to the linear order
of $\tau_{Wi}$: $a_{Wi}(t)\simeq a(t)\left[1-\tau_{Wi}(t)\right]$,
satisfying the trace condition ${\rm Tr}(a_{Wi})=3a(t)$ \citep[also
see][for the similar
discussion]{BondMyers:96,HuiBertschinger:96,IpSchmidt:16}.

As discussed in \citet{SherwinZaldarriaga:12} \citep[also
see][]{TakadaHu:13,Lietal:14a}, the super-survey modes affect the
clustering correlation function measured in the local
survey volume.
Extending the method in \citet{SherwinZaldarriaga:12} to include the
coherent tidal force, we can deduce that the clustering correlation
function of total matter, $\xi_W(\bbr)$, in the local volume is
modified, up to the linear order of the super-survey modes, as
\begin{eqnarray}
\xi_W(\bbr)&\equiv& \langle
 \delta(\bq_{W1})\delta(\bq_{W2})\rangle_{\bbr=\bq_{W1}-\bq_{W2}}\nonumber\\
 &=&
  \left(1+\frac{68}{21}\delta_\br\right) \xi\!\left(r^i+\frac{\delta_\br}{3}r^i+{\tau}_{Wij}r^j
\right)\nonumber\\
 &\simeq & \xi(r)+\left[\frac{68}{21}\xi(r)+\frac{1}{3}r^i\frac{\partial
		   \xi(r)}{\partial r^i}\right]\delta_\br
 + \frac{\partial \xi(r)}{\partial r^i} r^j
  {\tau}_{Wij}.
  \label{eq:xiW}
\end{eqnarray}
%
Note that the above correlation function is from the ensemble average of
sub-survey modes on a realization basis of the local volume that has the
fixed super-survey modes, $\delta_\br$ and $\tau_{Wij}$.
Eq.~(\ref{eq:xiW}) shows that, even if the real-space clustering is
isotropic, the correlation function measured in the local volume
generally becomes two-dimensional due to the coherent tidal force. It
causes an apparent anisotropic clustering in the local volume, and the
amount of the anisotropic clustering depends on angles between the
directions of $\tau_{Wij}$ and the separation vector $\bbr$.

Fourier-transforming Eq.~(\ref{eq:xiW}), we can find that the power
spectrum measured in the local volume is modified as
\begin{equation}
 P_W(\bk)\simeq
  P(k)+\delta_\br\left[\frac{47}{21}-\frac{1}{3}\frac{\partial \ln
		  P(k)}{\partial \ln k}\right]P(k)
   -\tau_{Wij}\hk_i\hk_j \frac{\partial P(k)}{\partial \ln k},
  \label{eq:pw}
\end{equation}
where $\hk_i\equiv k_i/k$.  Furthermore, in Appendix~\ref{app:ssc}, we
use the formulation in \citet{TakadaHu:13} to derive the full expression
for the responses of the power spectrum to the super-survey modes in the
weakly nonlinear regime. We show that the large-scale tide also causes a
change in the amplitude of the power spectrum. Thus the full expression
is given as
\begin{eqnarray}
 P_W(\bk)\simeq P(k)
  +\delta_\br\left[\frac{47}{21}-\frac{1}{3}\frac{\partial \ln P(k)}{\partial \ln
    k}\right]P(k)
  +\hk_i\hk_j\tau_{Wij}\left[\frac{8}{7}-\frac{\partial \ln P(k)}{\partial \ln k}\right]P(k).
\end{eqnarray}
The term with prefactor $8/7$ gives the effect of the large-scale tide
on the power spectrum amplitude.  For an arbitrary line-of-sight
direction that an observer takes, the anisotropic power spectrum in the
above equation appears exactly similar to the Alcock-Paczynski (AP)
distortion effect \citep{AlcockPaczynski:79} \citep[also
see][]{SeoEisenstein:03,HuHaiman:03,Takadaetal:06} as well as the
redshift-space distortion (RSD) effect, the Kaiser effect
\cite{Kaiser:87}. The large-scale overdensity $\delta_\br$ alters the
power spectrum amplitude as well as causes an isotropic dilation effect
that is given by the term involving $\partial P(k)/\partial k$. Note
that the terms involving $\delta_\br$ reproduce the 2-halo term of
Eq.~(27) in \citet{Lietal:14a} \citep[also see][]{TakadaHu:13}.  On the
other hand, the coherent tidal force causes a homogeneous anisotropic
dilation in all three directions, perpendicular and parallel to the
line-of-sight direction, while the RSD effect causes a distortion of the
clustering along the line-of-sight direction.  In particular, the terms
involving the power spectrum derivative, $\partial P/\partial
k$, causes a shift of the BAO peak location compared to what the BAO
location should be in the global background \citep[also see][for the
effect of $\delta_\br$ on the BAO peak location]{SherwinZaldarriaga:12}.
Due to the tensor nature of $\tau_{Wij}$, the directional dependence of
$k^ik^j$ causes a quadratic anisotropy in the power spectrum. Thus the
coherent tidal force causes a systematic error when estimating the
Hubble expansion rate and the angular diameter distance from the
anisotropic clustering via the AP effect.

Next we consider effects of super-survey modes on the redshift-space
power spectrum of galaxies. Galaxies are biased tracers of the
underlying matter distribution in the large-scale structure. In this
paper, we assume that the number density fluctuation field of galaxies
is locally related to the matter density fluctuation field at the same
position via a linear bias parameter $b$: $\delta_{g}(\bx)=b\delta_{\rm
m}(\bx)$. As shown in \citet{HuKravtsov:03}, the mean number density of
galaxies in a finite-volume survey is modulated from the global mean by
$\delta_\br$
as
\begin{equation}
\bar{n}_{{\rm g}W}\simeq \bar{n}_{\rm g}\left[1+b\delta_\br\right].
\end{equation}
Then the two-point correlation function of the galaxies in a local
volume is estimated relative to the local mean density, $\xi_{{\rm
g}W}(\bbr)=\avrg{n_{\rm g}(\bx)n_{\rm g}(\bx+\bbr)}/\bar{n}_{{\rm
g}W}^2-1$.  As discussed in \citet{Lietal:14a} \citep[also
see][]{Baldaufetal:16}, the real-space  power spectrum of galaxies is modified by
super-survey modes as
\begin{equation}
P_{{\rm g}W}(k)\simeq b^2\left(1-2b\delta_\br\right)P_W(k).
\end{equation}
Combining this with the super-survey effects (Eq.~\ref{eq:pw}) and the
 Kaiser RSD effect, we can find that the redshift-space power spectrum
 of galaxies is given as
\begin{equation}
 P^S_{{\rm g}W}(\bk) = \left[1+\beta \mu^2\right]^2\left[
P_{\rm g}(k)  +
\delta_\br\left\{\frac{47}{21}-2b-\frac{1}{3}\frac{\partial \ln P_{\rm g}(k)}{\partial \ln
    k}\right\}P_{\rm g}(k)
+\hk_i\hk_j\tau_{Wij}\left\{\frac{8}{7}-\frac{\partial \ln P_{\rm g}(k)}{\partial
		      \ln k}\right\}P_{\rm g}(k)
						   \right],
 \label{eq:psw}
\end{equation}
where $P_{\rm g}(k)$ is the real-space power spectrum in the global
background, $\mu$ is the cosine angle between the line-of-sight
direction and the wavevector $\bk$, and $\beta\equiv (1/b)\mathrm{d}\ln
D/\mathrm{d}\ln a$.
In the above equation we simply assumed that the Kaiser RSD effect
causes an additional distortion of the galaxy distribution, and treated
the effect as a multiplicative factor to the real-space power spectrum
(see below for further discussion).  {Thus the redshift-space power
specrum in the presence of the super-survey effects have redshift-space
distortions up to $\mu^6$, in the weakly nonlinear regime.} In the
following we focus on the effect of $\tau_{Wij}$ on the redshift-space
spectrum, and ignore the effect of $\delta_\br$ (i.e. set
$\delta_\br=0$).

Now we consider the multipole power spectra that are useful spectra to
quantify the RSD effects. Without loss of generality, we can assume that
the $z$-axis direction in the local coordinates is along the
line-of-sight direction of an observer. Taking into account the fact
that the coherent tidal force also causes an anisotropic dilation even
in the $xy$-plane perpendicular to the line-of-sight direction, where
the redshift distortion effect is absent, we can define the multipole
power spectrum as
\begin{equation}
 P^S_{{\rm g}\ell W}(k)=(2\ell + 1)\int^{1}_{-1}\!\!\frac{\mathrm{d}\mu}{2}
  \int^{2\pi}_0\!\frac{\mathrm{d}\varphi}{2\pi}~
  P^S_{{\rm g}W}(\bk){\cal L}_{\ell}(\mu),
  \label{eq:psw_l}
\end{equation}
where $\varphi$ and $\mu$ are the angle and cosine angle between the
coordinate axes and the wavevector $\bk$, i.e. $\bk\equiv
k\left(\sqrt{1-\mu^2}\cos\varphi,\sqrt{1-\mu^2}\sin\varphi,\mu\right)$
and ${\cal L}_{\ell }(\mu)$ is the $\ell$-th order Legendre polynomial;
${\cal L}_0(\mu)=1$ and ${\cal L}_2(\mu)=(3\mu^2-1)/2$ that are relevant
for the following calculation.

The monopole power spectrum is found to be
\begin{eqnarray}
 P^S_{{\rm g}0W}(k)&=&\left(1+\frac{2\beta}{3}+\frac{\beta^2}{5}\right)P_{\rm
  g}(k) +
  \left(\frac{1}{3}+\frac{2\beta}{5}+\frac{\beta^2}{7}\right)\left(
   {\tau}_{W11}+{\tau}_{W22}+{\tau}_{W33}
									   \right)
  \left[\frac{8}{7}-\frac{\partial \ln P_{\rm g}(k)}{\partial \ln
   k}\right]P_{\rm g}(k)
  \nonumber\\
  &=&\left(1+\frac{2\beta}{3}+\frac{\beta^2}{5}\right)P_{\rm
  g}(k),
\end{eqnarray}
where we used ${\rm Tr}({\tau_{Wij}})=0$.  Thus the coherent tidal force
does not affect the monopole power spectrum {because of the trace-free nature of $\tau_{Wij}$}.

On the other hand, the super-survey tide causes a modulation in the
quadrupole power spectrum:
%
\begin{eqnarray}
P^S_{{\rm g}2W}(k)&=&\left(\frac{4\beta}{3}+\frac{4\beta^2}{7}\right)P_{\rm
 g}(k)
 +\left(1+\frac{22\beta}{21}+\frac{3\beta^2}{7}\right)
 {\tau}_{W33} \left[\frac{8}{7}-\frac{\partial \ln P_{\rm g}(k)}{\partial \ln
   k}\right]P_{\rm g}(k)
 \label{eq:p2_tau}
\end{eqnarray}
where we used the fact ${\tau}_{W11}+{\tau}_{W22}=-{\tau}_{W33}$ in
deriving the above equation. Since the quadrupole power spectrum
amplitude depends on $\beta$, or in other words no contribution from the
monopole power spectrum, it is a useful probe of the growth
rate. However, the coherent tidal force causes an extra contribution to
the quadrupole power spectrum (the 2nd term on the r.h.s.). As we
emphasized above, the tidal effect varies with a position of the survey
region, and in this sense ${\tau}_{Wij}$ is a statistical variable.
Note that, even if the coherent density mode $\delta_\br$ exists in the
survey region, it only affects the amplitude of the quadrupole power
spectrum, and the effect is therefore perfectly degenerate with the bias
parameter.

Similarly one can compute the
extra contribution to the higher-order multipole power spectra:
\begin{eqnarray}
P^S_{{\rm g}4W}(k)&=&\frac{8\beta^2}{35}P_{\rm
 g}(k)+\left(\frac{24\beta}{35}+\frac{136\beta^2}{385}\right){\tau}_{W33}
  \left[\frac{8}{7}-\frac{\partial \ln P_{\rm g}(k)}{\partial \ln
   k}\right]P_{\rm g}(k),
 \nonumber\\
 P^S_{{\rm g}6W}(k)&=&\frac{8\beta^2}{77}{\tau}_{W33}
   \left[\frac{8}{7}-\frac{\partial \ln P_{\rm g}(k)}{\partial \ln
   k}\right]P_{\rm g}(k)
\end{eqnarray}
and $P^S_{\ell W}(k)=0$ for $\ell\ge 8$.  Thus the coherent tidal force
generally induces a non-vanishing $P_6$ power spectrum, which is absent
in the Kaiser formula.

In the following, we consider $P_{{\rm g}2W}^S(k)$, the leading-order
anisotropic power spectrum, to study the impact of the coherent tidal force.

\subsection{Super-sample covariance}
\label{sub:ssc}

We have so far shown that the super-survey tidal force affects a measurement of
the redshift-space power spectrum, and here estimate how the effect is
important compared to a statistical precision of the power spectrum
measurement. 

Extending the formulation for the real-space power spectrum in
\citet{Scoccimarroetal:99} \citep[also
see][]{TakadaBridle:07,TakadaHu:13}, we can write down an estimator for
the quadrupole power spectrum in a given survey region:
\begin{equation}
\hat{P}^S_{{\rm g}2}(k_i)\equiv \frac{5}{V_W}\int_{|\bk|\in
 k_i}\!\frac{\mathrm{d}^3\bk}{V_{k_i}}\tdelta_{{\rm
 g}W}(\bk)\tdelta_{{\rm g}W}(-\bk){\cal
 L}_2(\mu),
 \label{eq:est_p2}
\end{equation}
where $\tdelta_{{\rm g}W}(\bk)$ is the density fluctuation field of
galaxies convolved with the survey window, the prefactor $5$ is from the
definition of multipole power spectrum (Eq.~\ref{eq:psw_l}), $(2l+1)$,
the integral is over a shell in $k$-space of width $\Delta k$ and volume
$V_{k_i}\simeq 4\pi k_i^2\Delta k$ for $\Delta k/k_i\ll 1$.
We have here employed the
continuous  limit of discrete Fourier transforms under the approximation
that the total volume {for the Fourier transform}
is much greater than the survey region 
(see
Ref.~\cite{TakadaBridle:07,Kayoetal:13} for a
pedagogical derivation of power spectrum estimator and the covariance
 based on the discrete Fourier decomposition).

Similarly to the formulation in \citet{Schaanetal:14}, we introduce the
ensemble average of sub-survey modes for a {\em fixed} coherent tidal force,
$\tau_{Wij}$, denoted as $\avrg{\hspace{1em}}_{\tau_W}$. When we focus
on wavenumber modes satisfying $k\gg 1/L$, the average of the estimator
(\ref{eq:est_p2}) is computed as
\begin{eqnarray}
 \avrg{\hat{P}^S_{{\rm g}2}(k_i)}_{\tau_W}
 &\equiv&
  \frac{5}{V_W}\int_{|\bk|\in
 k_i}\!\frac{\mathrm{d}^3\bk}{V_{k_i}}\avrg{\tdelta_{{\rm
 g}W}(\bk)\tdelta_{{\rm g}W}(-\bk)}_{\tau_{W}}{\cal
 L}_2(\mu)\nonumber\\
 &\simeq & \frac{5}{V_W}\int_{|\bk|\in
  k_i}\!\frac{\mathrm{d}^3\bk}{V_{k_i}}
P^S_{{\rm g}W}(\bk; \tau_W){\cal
L}_2(\mu)\int\!\!\frac{\mathrm{d}^3\bq}{(2\pi)^3}|\tW(\bq)|^2\nonumber\\
 &=&
5\int_{|\bk|\in
  k_i}\!\frac{4\pi k^2~\mathrm{d}k}{V_{k_i}}\int_{-1}^{1}\!\frac{\mathrm{d}\mu}{2}
P^S_{{\rm g}W}(\bk;\tau_W){\cal
L}_2(\mu)\nonumber\\
 &=&  \int_{|\bk|\in
  k_i}\!\frac{4\pi k^2~\mathrm{d}k}{V_{k_i}}
  \left[
  \left(\frac{4\beta}{3}+\frac{4\beta^2}{7}\right)P_{\rm
 g}(k)
 +\left(1+\frac{22\beta}{21}+\frac{3\beta^2}{7}\right)
 {\tau}_{W33}
   \left\{\frac{8}{7}-\frac{\partial \ln P_{\rm g}(k)}{\partial \ln
   k}\right\}P_{\rm g}(k)\right]
  \nonumber\\
 &\simeq & 
  \left(\frac{4\beta}{3}+\frac{4\beta^2}{7}\right)P_{\rm
 g}(k_i)
 +\left(1+\frac{22\beta}{21}+\frac{3\beta^2}{7}\right)
 {\tau}_{W33}
  \left[\frac{8}{7}-\frac{\partial \ln P_{\rm g}(k)}{\partial \ln
   k}\right]_{k_i}P_{\rm g}(k_i)
\end{eqnarray}
where $P^S_{{\rm g}W}(\bk;\tau_W)$ is the power spectrum obtained by
setting $\delta_\br=0$ in Eq.~(\ref{eq:psw}).  Furthermore, because
the convolution changes the power spectrum only around $k \lesssim 1/L$
due to the nature of the window function, here we are
interested in the spectra of modes satisfying
$k \gg 1/L$. That is, we have used
$P^S_W(\bk-\bq;\tau_W)\simeq P^S_W(\bk;\tau_W)$ over the integral range of
$\mathrm{d}^3\bq$, and assumed that the power spectrum $P_{\rm g}(k)$ is
not a rapidly varying function within a $k$-bin. Thus the average of the
estimator (Eq.~\ref{eq:est_p2}) for a fixed
$\tau_{Wij}$ recovers Eq.~(\ref{eq:p2_tau}).

Now we introduce the ensemble average that is the average of the
estimator with varying positions of the survey regions, denoted
as $\avrg{\hspace{1em}}$:
%
\begin{eqnarray}
 \avrg{\hat{P}^S_{{\rm g}2}(k_i)}&\simeq&
  \left(\frac{4\beta}{3}+\frac{4\beta^2}{7}\right)P_{\rm
 g}(k_i)
 +\left(1+\frac{22\beta}{21}+\frac{3\beta^2}{7}\right)
 \avrg{{\tau}_{W33}}
   \left[\frac{8}{7}-\frac{\partial \ln P_{\rm g}(k)}{\partial \ln
   k}\right]_{k_i}P_{\rm g}(k_i)\nonumber\\
   &\simeq&
   \left(\frac{4\beta}{3}+\frac{4\beta^2}{7}\right)P_{\rm
 g}(k_i),
\end{eqnarray} 
where
we assumed $\avrg{\hat{\tau}_{W33}}=0$, i.e. the
average of the coherent tidal force is vanishing in the ensemble average
sense.
Thus the ensemble average of the estimator (\ref{eq:est_p2}) recovers
the quadrupole power spectrum in the Kaiser formula.

Now we consider the covariance of the quadrupole power spectrum, defined
in terms of the estimator as
\begin{equation}
 C_{ij}\equiv \avrg{\hat{P}_2(k_i)\hat{P}_2(k_j)}-
  \avrg{\hat{P}_2(k_i)}
  \avrg{\hat{P}_2(k_j)}
\end{equation}
Similarly to \citet{TakadaHu:13}, we find that the covariance is
decomposed into two contributions
\begin{equation}
 {\bf C}\simeq  {\bf C}^{\rm G}+{\bf C}^{\rm SSC}. 
\end{equation}
The first term is a Gaussian term, and the second term is the
non-Gaussian error arising from the coherent tidal force on which we
focus in this paper. Here we ignored the trispectrum contribution of
sub-survey modes to the sample variance for simplicity.

Following method in \citet{Guziketal:10} and \citet{TakadaHu:13}, we can
compute the Gaussian term as
\begin{eqnarray}
 C_{ij}^{\rm G}&\simeq& {\delta_{ij}^K}\frac{25}{V_W}\frac{(2\pi)^3}{V_{k_i}}
  \int_{|\bk|\in
  k_i}\!\frac{\mathrm{d}^3\bk}{V_{k_i}} 2 \left[1+\beta \mu^2\right]^2
  \left[P_{\rm g}(k)+\frac{1}{\bar{n}_{\rm g}}\right]^2
  \left[{\cal L}_2(\mu)\right]^2\nonumber\\
 & =&
  {\delta_{ij}^K}\frac{50}{V_W}\frac{(2\pi)^3}{V_{k_i}}\int_{|\bk|\in
  k_i}\!\frac{4\pi k^2 \mathrm{d}k}{V_{k_i}}
  \int_{-1}^{1}\!\frac{\mathrm{d}\mu}{2}\left[1+\beta \mu^2\right]^2
  \left[P_{\rm g}(k)+\frac{1}{\bar{n}_{\rm g}}\right]^2
  \left[{\cal L}_2(\mu)\right]^2\nonumber\\
 &\simeq & \delta^K_{ij}\frac{(2\pi)^3}{V_W V_{k_i}}10
  \left[1+\frac{44}{21}\beta+\frac{18}{7}\beta^2
   +\frac{340}{231}\beta^2+\frac{415}{1287}\beta^4
  \right]\left[P_{\rm g}(k)+\frac{1}{\bar{n}_{\rm g}}\right]^2,
\label{eq:cov_g}
\end{eqnarray}
where we have included the shot noise term arising from a finite number
of sampled galaxies, given by the terms including $1/\bar{n}_{\rm g}$.
The Gaussian covariance scales as $1/V_{W}$. More exactly speaking, it
scales as the number of independent $k$-modes in the shell as
\begin{equation}
 N_{\rm mode}(k_i)=\frac{V_{k_i}V_W}{(2\pi)^3}\simeq \frac{4\pi k_i^2\Delta k~ V_W}{(2\pi)^3}
\end{equation}
The Gaussian covariance matrix is diagonal, and in other words, its
off-diagonal components are vanishing. 

On the other hand, the super sample covariance (SSC) term is given as
\begin{eqnarray}
 C^{\rm SSC}_{ij}=\left(1+\frac{22\beta}{21}+\frac{3\beta^2}{7}\right)^2
    \left[\frac{8}{7}-\frac{\partial \ln P_{\rm g}(k)}{\partial \ln
   k}\right]_{k_i}
   \left[\frac{8}{7}-\frac{\partial \ln P_{\rm g}(k)}{\partial \ln
   k}\right]_{k_j}P_{\rm g}(k_i)P_{\rm g}(k_j)
   \sigma_{\tau33}^2.
  \label{eq:cov_ssc}
\end{eqnarray}
where $\sigma_{\tau 33}^2$ can be calculated using
Eq.~(\ref{eq:sigma_tau}) for a given cosmological model and survey
window, and we have assumed $\avrg{\delta_\br\tau_{W33}}\ll
\sigma_{\tau 33}^2$ for a reasonable window.
The SSC covariance has off-diagonal components. 

In the following we will use Eqs.~(\ref{eq:cov_g}) and
(\ref{eq:cov_ssc}) to compute the covariance for a measurement of the
quadrupole power spectrum for a hypothetical galaxy survey.

\section{Results}
\label{sec:results}

We throughout this paper employ cosmological parameters that are
consistent with the nine-year WMAP results \citep{WMAP9}: $\Omega_{\rm
c0}h^2=0.1165$, $\Omega_{\rm b0}h^2=0.02248$, and
$\Omega_{\Lambda}=0.7055$ for the density parameters of CDM, baryon and
the cosmological constant, $A_s=2.455\times 10^{-9}$ for the amplitude
of the primordial curvature perturbation, $n_s=0.967$ for the tilt of
primordial power spectrum, and $h=0.687$ for the Hubble constant. In
this model $\sigma_8=0.815$, which is the variance of present-day,
linear matter fluctuations within a sphere of radius $8~{\rm Mpc}/h$.


\begin{figure}[tb]
    \centering \includegraphics[width=8.5cm]{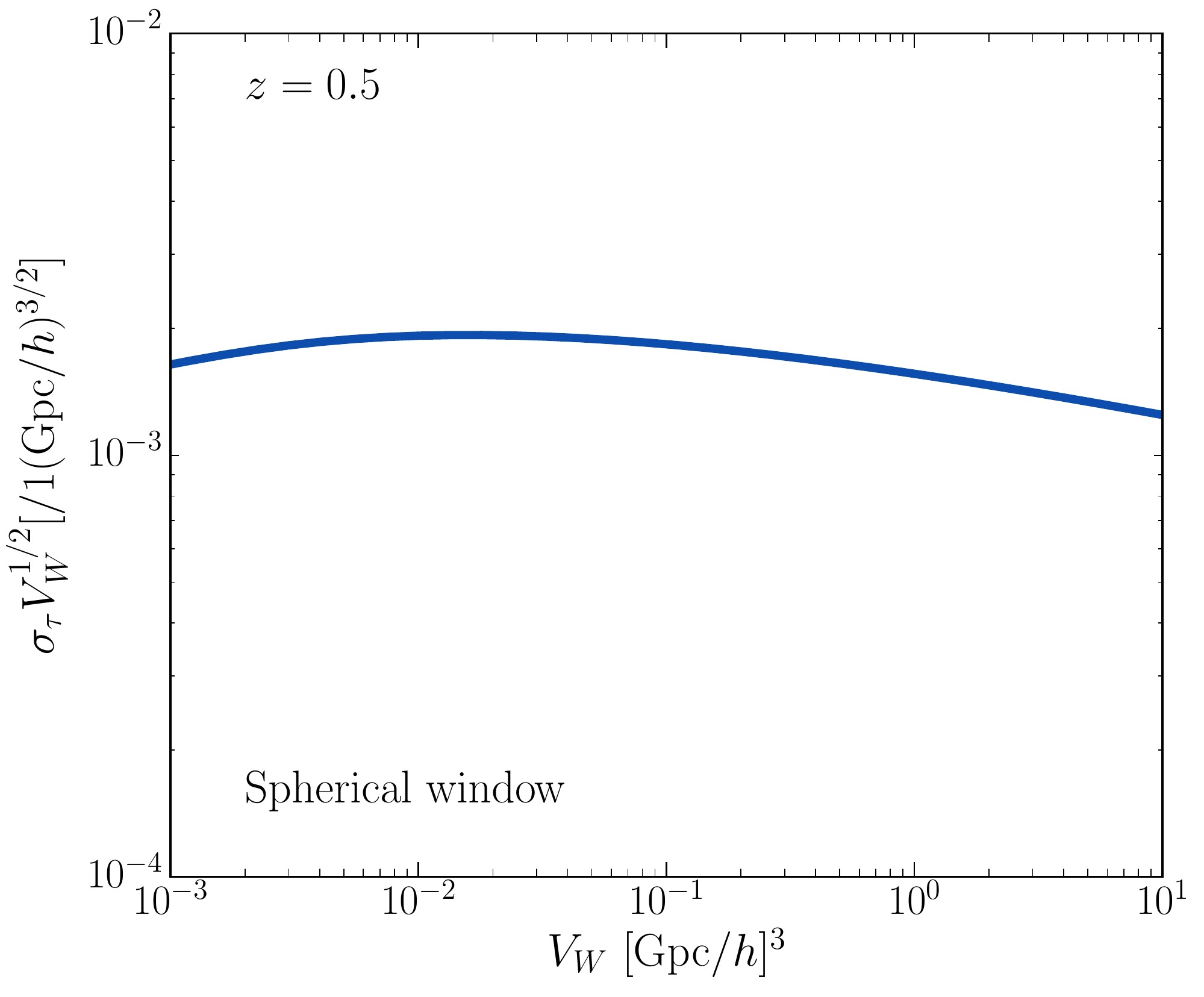}
 \includegraphics[width=8.5cm]{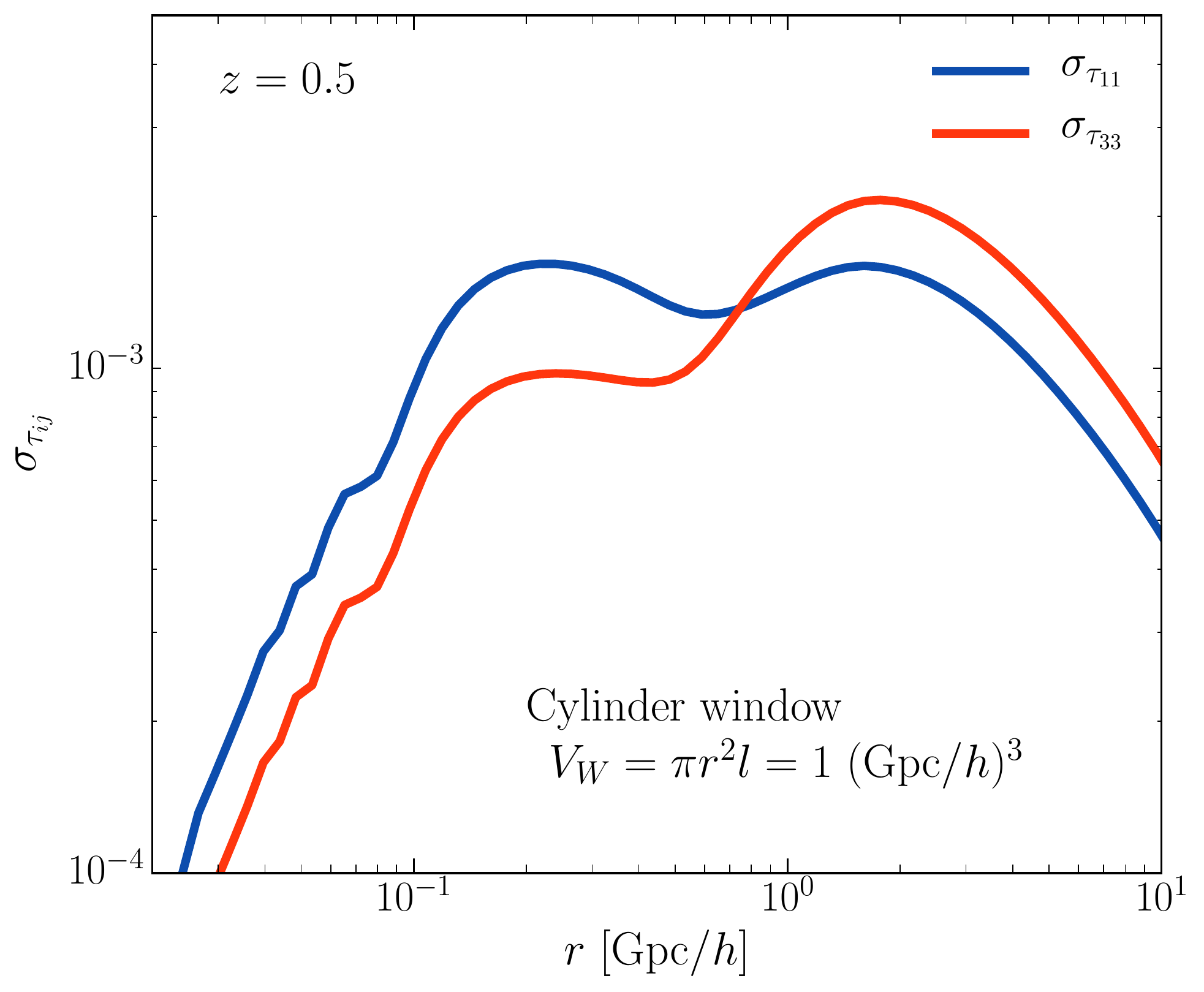}
 \caption{The rms of linear gravitational tidal field {convolved} with the
    survey window, $\sigma_{\tau ij}$ for a $\Lambda$CDM model and
    $z=0.5$ (see Eqs.~\ref{eq:sigma_tau} and \ref{eq:sigma_tau_iso}).
    {\em Left panel}: the rms as a function of survey volume for
    spherical window, $V_W=4\pi r^3/3$. In this case, the tidal tensor
    becomes diagonal; $\sigma_{\tau}\equiv \sigma_{\tau 11}=\sigma_{\tau
    22}=\sigma_{\tau 33}$. {\em Right panel}: the rms for cylinder
 windows of fixed volume $V_W=\pi r^2 \ell = 1~({\rm Gpc}/h)^3$, as a function of the
 radius of base circle, $r$. Here we assume that the height of cylinder
 is along the 3-axis (the light-of-sight) direction, and the base circle
 is in the plane perpendicular to the line-of-sight direction (therefore
 $\sigma_{\tau11}=\sigma_{\tau22}$). For an elongated cylinder window,
 i.e. a tube-like shaped survey, $\sigma_{\tau 11}$ has a greater amplitude, while
 $\sigma_{\tau 33}$ has a greater amplitude for a pill-like shape. When
 $r\sim \ell$, $\sigma_{\tau11}\simeq \sigma_{\tau33}$, and the linear
 variance has a largest amplitude. 
 } \label{fig:tau}
\end{figure}
The key quantity to characterize the effect of coherent tidal force is
the variance of linear tidal field averaged over the survey window,
$\sigma_\tau$ (Eq.~\ref{eq:sigma_tau}).  The left
panel of Fig.~\ref{fig:tau} shows the variance for a $\Lambda$CDM model,
for spherical window as a function of survey volume $V_W$. Other
covariance term scales with $1/V_W$, so the curve shows the relative
contribution of the coherent tidal force to the sample
variance. Likewise the effect of super-survey overdensity
$\sigma_\br$ \citep{TakadaHu:13}, the super-survey covariance has a
weak dependence on the volume. For a sufficiently large cosmological
volume such as $V_W\gtrsim 1~({\rm Gpc}/h)^3$, $\sigma_\tau\sim 10^{-3}$.

As can be found from Eq.~(\ref{eq:sigma_tau}), the different components
of the linear variance of super-survey tidal tensor, $\sigma_{\tau ij}$,
depends on the shape of survey window. The right panel of
Fig.~\ref{fig:tau} studies this for a cylinder window as a function of
the different shape, for a fixed survey volume of $V_W=\pi
r^2\ell=1~({\rm Gpc}/h)^3$. {When $r\ll \ell$, a survey window
corresponds to a survey being ``narrow'' in area coverage on the sky,
but deep in redshift direction -- a ``tube-shaped'' survey. A
survey with $\ell \ll r$ corresponds to a survey being ``wide'' in area, but shallow
in redshift -- a ``pill-shaped'' survey.}
The linear variance components have different amplitudes
depending on angles between the coordinate axes and the principal axes
of tidal tensor.  Here we consider the line-of-sight direction to lie
along the 3rd-axis direction of an observer coordinate's system and the
height direction of the cylinder window ($\ell$-direction); in this case
$\sigma_{\tau 11}=\sigma_{\tau {22}}$.  The variance components,
$\sigma_{\tau 33}\simeq \sigma_{\tau 11}$ when $r\simeq 0.7~{{\rm
Gpc}/h}$ or equivalently $\ell \simeq r$. For either case of extreme
``tube'' or ``pill'' shape, 
one component 
$\sigma_{\tau 11}$ or $\sigma_{\tau 33}$ has a greater
amplitude than the other. However, the variance amplitude gets smaller
due to cancellation effect of the linear variances \citep[also
see][]{TakadaHu:13}. However, the extreme cases are not desirable,
because one length scale of the volume can be in the nonlinear regime,
and the linear-order approximation of the super-survey modes breaks
down.

\begin{figure}
    \centering \includegraphics[width=8.5cm]{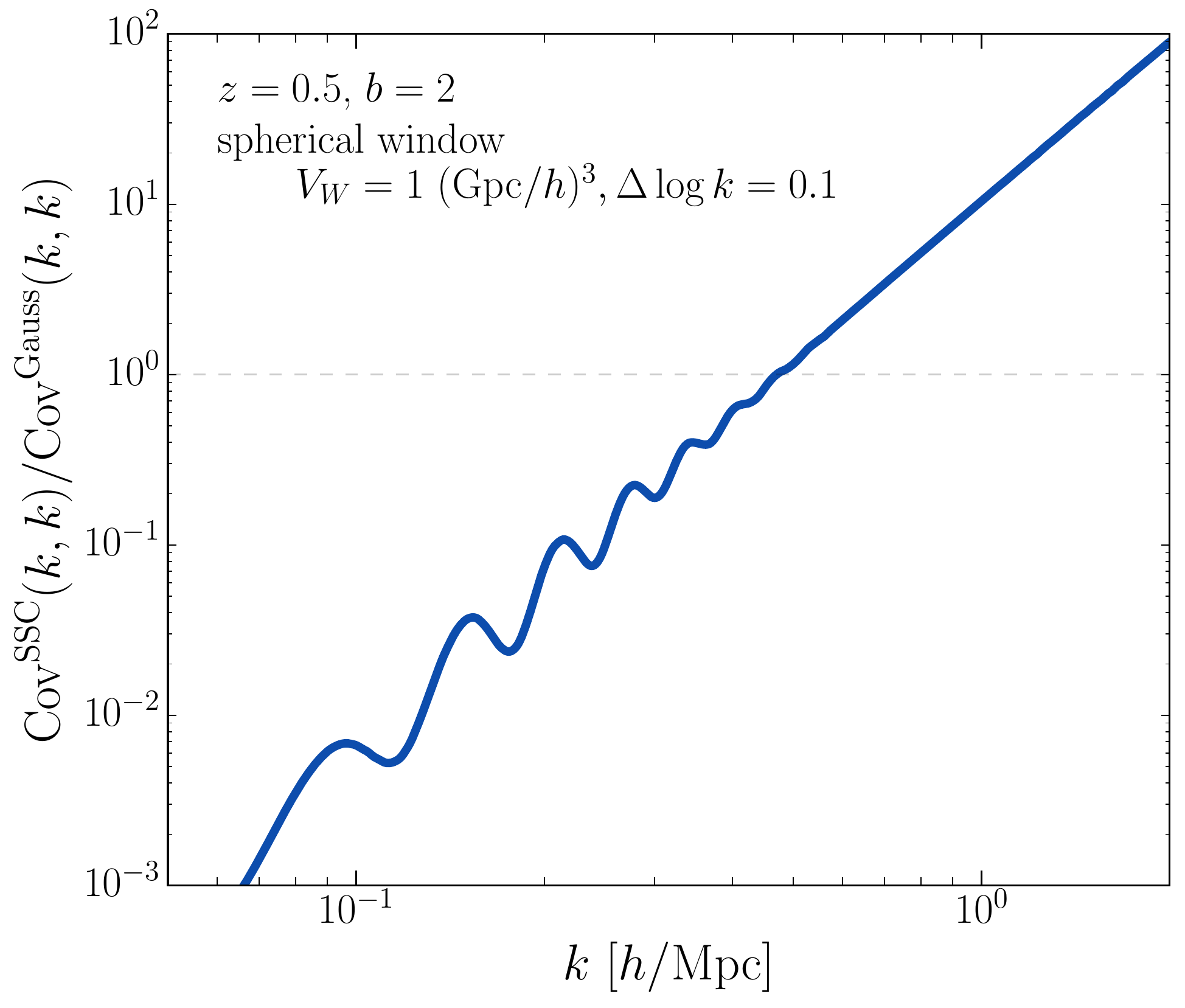}
 \caption{Shown is how the super-survey tidal force causes an increase
 in the sample variance in a measurement of the quadrupole power
 spectrum of redshift-space galaxy distribution. The curve shows the
 super-sample covariance (SSC) contribution to the sample variance
 relative to the Gaussian variance, for a survey with volume $1~({\rm
 Gpc}/h)^3$, at $z=0.5$ and galaxies with linear bias $b=2$
 (Eqs.~\ref{eq:cov_g} and \ref{eq:cov_ssc}). Here we ignored the shot
 noise contribution due to a finite number density of galaxies. Since
 the Gaussian term depends on the bin width of wavenumber, we assumed a
 binning of $\Delta \log k=0.1$ (10 bins in one decade of
 wavenumber). The SSC effects causes a significant sample variance at
 $k\gtrsim 0.5~h/{\rm Mpc}$.}  \label{fig:cov}
\end{figure}
Fig.~\ref{fig:cov} compares the Gaussian and super-survey covariance
terms in the covariance matrix of the quadrupole power spectrum, for a
spherical window of $V_W=1~({\rm Gpc}/h)^3$ (see Eqs.~\ref{eq:cov_g} and
\ref{eq:cov_ssc}). Here we assume a survey probing the three-dimensional
distribution of galaxies at $z=0.5$ and with linear bias parameter
$b=2$, which roughly resemble SDSS CMASS-type galaxies
\citep{Reidetal:12}. Here we ignored the effect of a finite number
density of the galaxies. Since the Gaussian covariance depends on the
bin width of wavenumber, we employ $\Delta \log k=0.1$. Note that, in
order to show the effect of the coherent tidal force on the BAO
features, we employed a much finer $k$-binning to plot the curve, but
used $\Delta\log k=0.1$ to compute the Gaussian term at each
$k$-bin. The figure shows that the super-survey effect gives a dominant
contribution to the sample variance in the weakly nonlinear regime,
$k\gtrsim 0.7~{h/{\rm Mpc}}$.
 
\begin{figure}
    \centering \includegraphics[width=8.5cm]{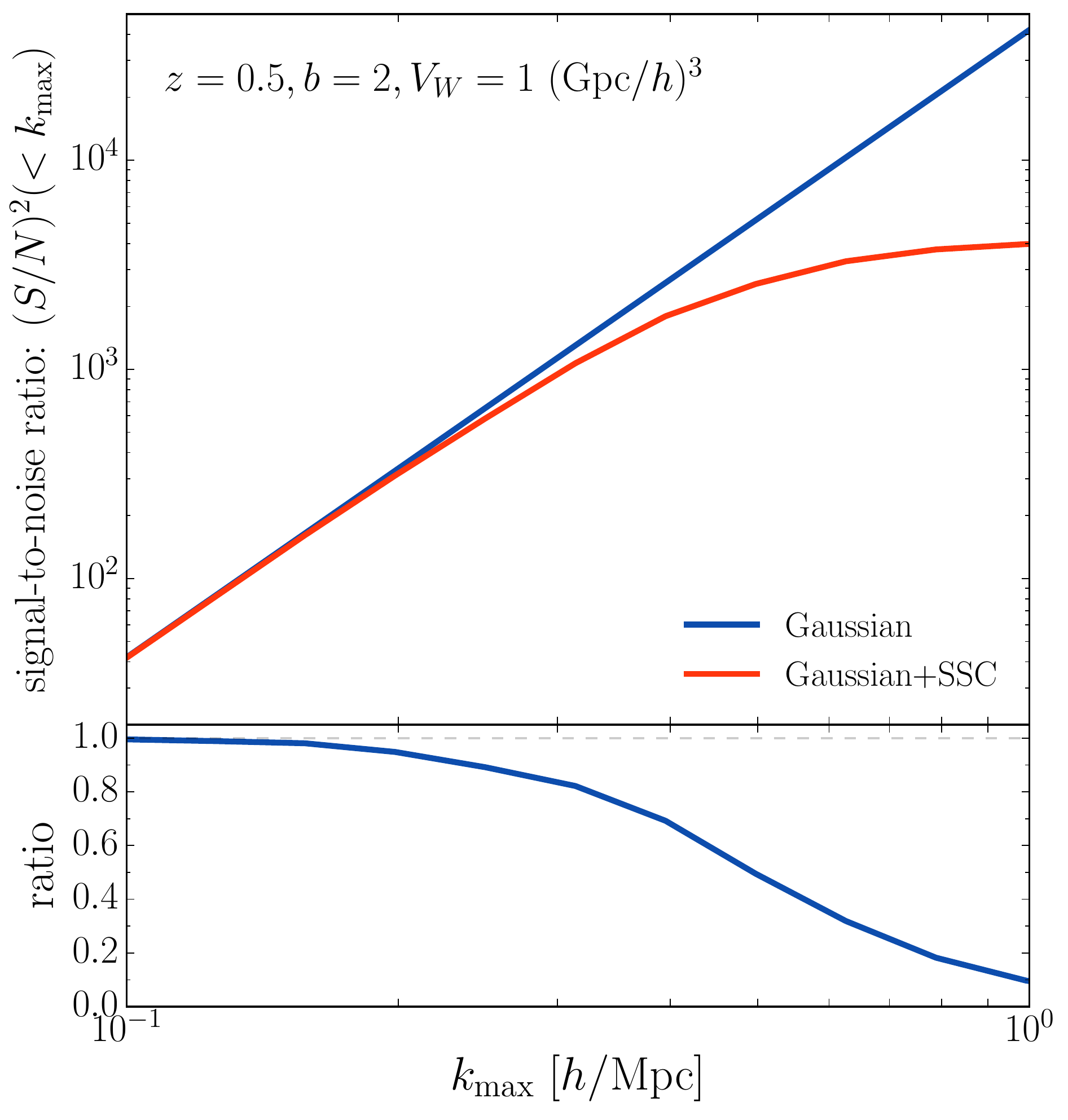}
    \centering \includegraphics[width=8.5cm]{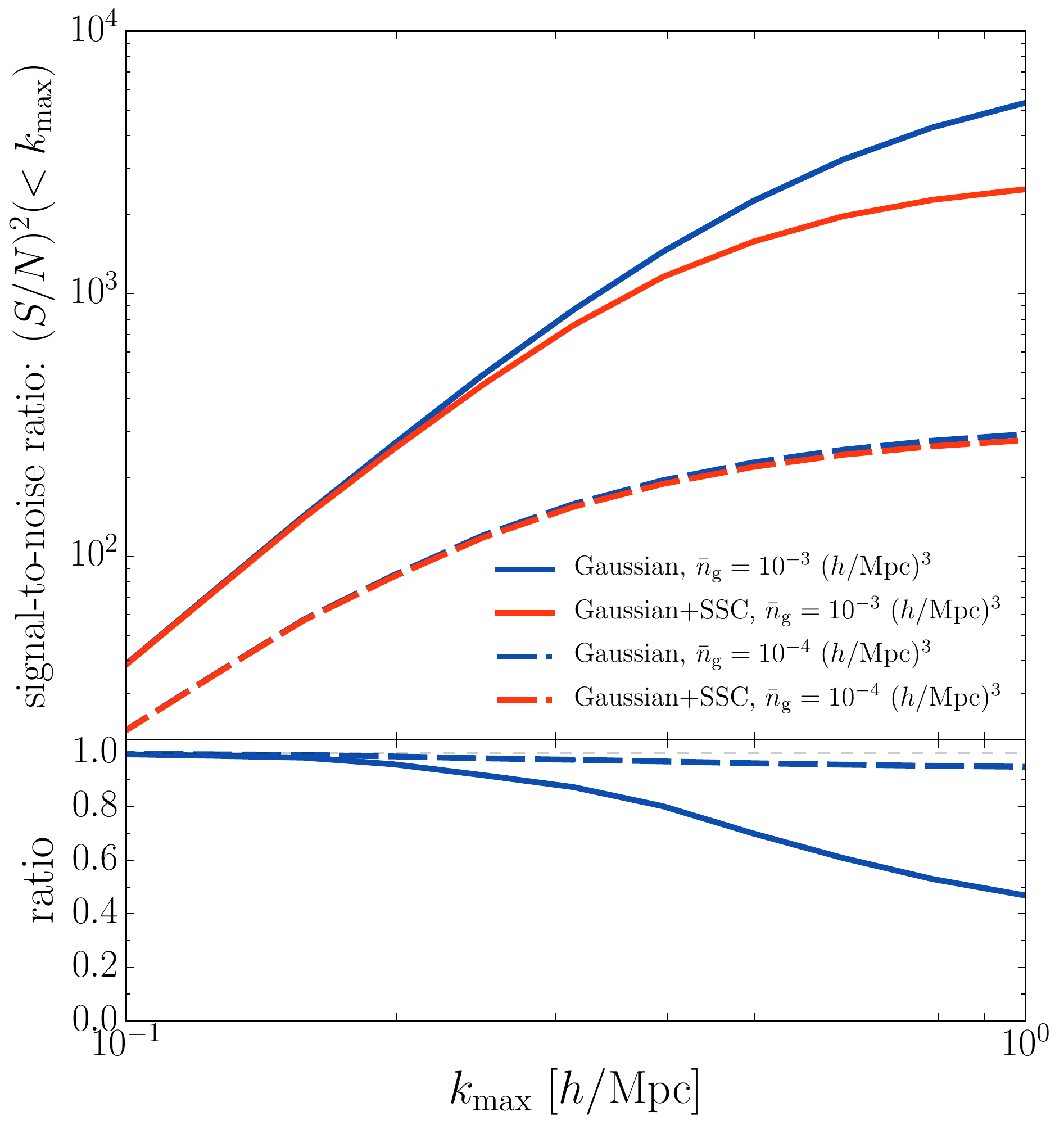}
 \caption{Cumulative signal-to-noise ratio (S/N) for a measurement of
 the quadrupole power spectrum, as a function of maximum wavenumber
 $k_{\rm max}$. We here assumed $V_W=1~({\rm Gpc}/h)^3, z=0.5$ and $b=2$
 as in the previous figure. S/N does not depend on the bin width of $k$.
 {\it Left panel}: The results for an infinite number density of
 galaxies; we ignored the shot noise contribution. The upper or lower
 curves in the upper plot are the results when assuming the Gaussian
 covariance or including the SSC contribution, respectively. The lower
 panel shows the ratio.  {\it Right panel}: The similar plot, but for a
 finite number density of galaxies: $\bar{n}_{\rm g}=10^{-3}$ or
 $10^{-4}~(h/{\rm Mpc})^3$, respectively.} \label{fig:SN}
\end{figure}
As one demonstration of the impact of the coherent tidal force on a
measurement of the quadrupole power spectrum in redshift space, we study
a cumulative signal-to-noise ratio, defined as
\begin{equation}
 \left(\frac{S}{N}\right)^2_{\le k_{\rm max}}\equiv
  \sum_{k_i k_j \in k_{\rm max}}P^S_{{\rm g}2W}(k_i)\left[{\bf
						     C}^{-1}\right]_{ij}
  P^S_{{\rm g}2W}(k_j),
  \label{eq:sn}
\end{equation}
where ${\bf C}^{-1}$ is the inverse of the covariance matrix, and the
summation runs over all wavenumber bins up to a given maximum wavenumber
$k_{\rm max}$. This quantity does not depend on the bin width. The
inverse of $(S/N)$ gives a statistical precision of measuring the
overall amplitude of the power spectrum, if the shape is completely
known.  Fig.~\ref{fig:SN} shows the results. The super-survey tidal
force causes a degradation in the power spectrum measurement, at $k_{\rm
max}\gtrsim \mbox{a few}~h/{\rm Mpc}$, and for a galaxy survey with a
high number density such as $\bar{n}_{\rm g}\simeq 10^{-3}~(h/{\rm
Mpc})^3$, which is the case for the WFIRST-AFTA survey
\citep{WFIRST:15}.

\section{Discussion}
\label{sec:discussion}

We have derived a formula to describe the effect of super-survey,
coherent tidal force on the redshift-space power spectrum measured in a
finite volume survey. The large-scale coherent overdensity and
tidal field both arise from the Hessian of the
long-wavelength gravitational potential, and are of equal importance.
Since the super-survey modes are not direct
observables, the effects on cosmological observables need to be
theoretically modeled.
The super-survey tide causes a characteristic, anisotropic
clustering pattern in the distribution of the tracers,
in all three directions perpendicular and parallel to the line-of-sight
direction (see Eq.~\ref{eq:psw}). This effect appears to be exactly
similar to the geometrical Alcock-Paczynski (AP) distortion as well as
the redshift-space distortion effect of peculiar velocities. We then
derived a formula to model the contribution of the coherent tidal force
to the sample variance in a measurement of the quadrupole redshift-space
power spectrum. We showed that the super-sample variance is not
negligible if including the power spectrum information up to the weakly
nonlinear regime, $k\gtrsim \mbox{a few}~h/{\rm Mpc}$ or for a galaxy
survey with a high number density such as $\bar{n}_{\rm g}\simeq
10^{-3}~(h/{\rm Mpc})^3$. In our derivation, we have not yet properly
included the super-survey effects on the nonlinear Kaiser factor. This
can be done by using the perturbation theory, and is our future work.

For a galaxy survey with a volume coverage greater than $\sim ({\rm
Gpc/h})^3$ and $z=0.5$ as in the SDSS survey, the linear variance of
super-survey tidal force $\sigma_{\tau}\sim 10^{-3}$ for a $\Lambda$CDM
model. This implies that the super-survey tide causes about 0.1\%
anisotropy in the clustering distribution. However, the expectation
value of the variance
is after the angle average, compared to the variance of coherent density
contrast: $\sigma_\tau\simeq \sqrt{4/45}\sigma_\br\simeq \sigma_\br/3.4$
(see Eq.~\ref{eq:sigma_tau_iso}).  If the principal axes of the
super-survey tidal tensor have an alignment to directions parallel
and/or perpendicular to the line-of-sight direction, the tide could have
a similar amplitude as $\delta_\br$ in a particular realization:
$\tau\sim \delta_\br$ corresponding to $\sim 0.3\%$ anisotropy. Since
the current state-of-the-art SDSS BOSS survey already achieved about 1\%
accuracy for the BAO distance measurements \citep{Alametal:16}, the
super-survey tide could cause a bias by an amount of the $1\sigma$
statistical error, if the SDSS survey volume is embedded into a
particular region where the aligned tide has a $3\sigma$ value.
Hence, it would be more important
to study how the coherent tidal force could cause a bias in measurements
of the cosmological distances via the AP test as well as the growth rate
via the RSD effect. This requires to propagate the expected statistical
accuracy of the redshift-space power spectrum measurement into parameter
estimation for a given survey geometry, including marginalization over
other parameters.  This is our future work, and will be presented
elsewhere.


The redshift-space clustering of galaxies is anisotropic by nature, and
the coherent tidal force causes a similar anisotropic clustering pattern
in the observed distribution. For the monopole power spectrum such as
the weak lensing power spectrum, the effect disappears at the first
order of $\tau_{Wij}$ due to the trace-less nature ${\rm
Tr}(\tau_{Wij})=0$. There are other effects of the coherent tidal force
that can be observed in principle from upcoming wide-area galaxy
surveys. First, it is shown that the coherent tidal force causes a
modification of dark matter halo formation via a coupling of the inertia
of mass distribution in a proto-halo region with the coherent tidal
force, leaving a non-local bias effect relative to the underlying matter
distribution at the second order: $\partial \delta_{\rm h}/\partial
(\tau_W^2)=b_\tau $ \citep{Chanetal:12,Saitoetal:14}. The non-local bias
can be measured by combining measurements of the power spectrum
(two-point) and bispectrum (three-point) of large-scale structure
tracers. Another observable is the correlation of the large-scale tidal
force with shapes of galaxies at much smaller scales, the so-called
intrinsic alignments \citep{Heavensetal:00,HirataSeljak:04}. The
intrinsic alignments are one of the major systematic effects for ongoing
and upcoming weak lensing surveys. Conversely, the intrinsic alignments
can be regarded as a ``signal'', rather than a contaminating systematic
error, and can be measured from these wide-area galaxy surveys in order
to constrain the large-scale tidal force
\cite{Schmidtetal:15,Chisarietal:16}.  Furthermore, a better
understanding of the nonlinear mode coupling allows one to use a
combination of the observed sub-survey modes to estimate the large-scale
tidal field \citep{Penetal:12,Zhuetal:16a,Zhuetal:16}.

In order to realize the effect of coherent tidal force on structure
formation in the deeply nonlinear regime, such as halo formation, we
need to use $N$-body simulations. For this purpose, a separate universe
simulation technique would be powerful; since the large-scale tidal
force can be absorbed into the perturbed scale factors along each
coordinate axis, $a_W(t)\simeq a(t)\left[1-\tau_{Wi}\right] $, we can follow
the full nonlinear mode coupling of the large-scale tide with sub-box
modes by running $N$-body simulations in the perturbed background.  For
the coherent overdensity $\delta_\br$, the effect for a
$\Lambda$CDM model can be absorbed as an apparent curvature, even if the
global background is flat. Several works have developed the separate
universe simulation technique to study the mode coupling effect of
$\delta_\br$ with sub-box modes
\citep{Lietal:14a,Lietal:14b,Wagneretal:15,Baldaufetal:16,Lazeyrasetal:16,Lietal:16}.
The separate universe simulation allows for a better calibration of
various effects such as the super-sample covariance and the local halo
bias, without running a large number of huge box simulations.
In a very similar way we believe that the separate universe simulation
technique can be applied to the large-scale tidal effect.  Recently
\citet{IpSchmidt:16} developed a unified formula to model the effect of
the coherent tidal force on the evolution of sub-survey modes within the
framework of general relativity. However, there are in general two
contributions to the large-scale tidal field: the internal tidal force
arising from the anisotropic matter distribution within a finite-volume
boundary and the external tidal force that is not specified by the
internal boundary conditions \citep[also
see][]{BondMyers:96,HuiBertschinger:96}. Nevertheless, as long as we are
interested in the effects of the linear tidal force, it would be
possible to develop a separate universe simulation technique to include
the large-scale tide in the simulation as well as to study the effect on
nonlinear structure formation. If this is true, the separate universe
simulations would give us a better way to calibrate the large-scale
tidal effects on various cosmological observables.  This is in progress
and will be presented elsewhere.\\

\smallskip{\em Acknowledgments.--} We thank Matias Zaldarriaga, Fabian
Schmidt, Atsushi Taruya and Tobias Baldauf for useful discussion, and we
also thank to YITP, Kyoto University for their warm hospitality. KA is
supported by the Advanced Leading Graduate Course for Photon Science at
the University of Tokyo. MT is supported by World Premier International
Research Center Initiative (WPI Initiative), MEXT, Japan, by the FIRST
program ``Subaru Measurements of Images and Redshifts (SuMIRe)'', CSTP,
Japan. MT is supported by Grant-in-Aid for Scientific Research from the
JSPS Promotion of Science (No.~23340061, 26610058, and 15H05893), MEXT Grant-in-Aid
for Scientific Research on Innovative Areas (No.~15K21733,
15H05892) and by JSPS Program for Advancing Strategic International
Networks to Accelerate the Circulation of Talented Researchers.

\appendix
\section{Takada \& Hu derivation of power
 spectrum response to super-survey modes}
 \label{app:ssc}


In this appendix we derive the responses of the real-space power
spectrum to the long-wavelength tidal force, based on the formulation in
\citet{TakadaHu:13}. 

Taking into account the survey window, The observed field of the matter
fluctuation field can be defined as
%
\begin{equation}
 \delta_W(\bx)=\delta(\bx)W(\bx),
\end{equation}
whose Fourier transform is a convolution
\begin{equation}
\tdelta_W(\bk) = \int\!\! \frac{d^3 \bq}{(2\pi)^3}~ \tilde{W}(\bq)
 \tdelta(\bk-\bq).
\end{equation}
%

In order to study how the large-scale tide causes an anisotropic
modulation in the measured power spectrum, let us define an estimator of
the {\it two-dimensional} power spectrum of wavevector $\bk$ as
\begin{equation}
\hat{P}(\bk)=\frac{1}{V_W}\tdelta_W(\bk)\tdelta_W(-\bk),
\label{eq:ps2d_est}
\end{equation}
Note that the wavevector bin $\bk$ can be finite, compared to the
fundamental mode of a survey, $k_f\simeq 2\pi/L$, and in that case the
above estimator is defined from a sum of the modes within the bin
width.  The power spectrum estimator satisfies a parity
invariance:
\begin{equation}
 \hat{P}(\bk)=\hat{P}(-\bk).
\end{equation}
The ensemble average of the estimator is found to recover the
underlying true power spectrum
\begin{eqnarray}
 \avrg{\hat{P}(\bk)}&=&\frac{1}{V_W}\int\!\!\frac{\mathrm{d}^3\bq}{(2\pi)^3}~|\tW(\bq)|^2P(\bk-\bq)
  \simeq
  P(\bk)\frac{1}{V_W}\int\!\!\frac{\mathrm{d}^3\bq}{(2\pi)^3}~|\tW(\bq)|^2=P(\bk)
  \label{eq:Pk}
\end{eqnarray}
Here we have used that $P(\bk-\bq)\simeq P(\bk)$ over the integration range
of $d^3\bq$ which the window function supports and also assumed that $P(\bk)$ is
not a rapidly varying function within the 
$k$-bin.  In the third
equality on the r.h.s., we have used the general identity for the window
{function}: 
\begin{equation}
V_{W}=\int\!d^3\bx~W^n(\bx)=
\int\!\left[\prod_{a=1}^n \frac{d^3\bq_a}{(2\pi)^3}
\tW(\bq_a)\right](2\pi)^3\delta^3_D(\bq_{1\ldots n}),
\label{eq:generalVW}
\end{equation}
where $\bq_{1 \ldots n}= \bq_1 +\ldots \bq_n$ here and below.
For $n=2$,
 $V_W=\int  |\tW(\bq)|^2 d^3\bq /(2\pi)^3$.

As we have discussed, the super-survey modes affect the power spectrum
measured in a finite-volume survey region. Hence, when a given survey
volume has super-survey modes of $\delta_\br$ and $\tau_{Wij}$, the
effects on power spectrum measured in the survey realization are
expressed as
\begin{equation}
P(\bk;\delta_\br,\tau_W)\simeq P(k)+\frac{\partial P(k)}{\partial 
 \delta_\br}\delta_\br + \frac{\partial P(k)}{\partial
 \tau_{Wij}}\tau_{Wij}.
 \label{eq:ps_resp}
\end{equation}
Here $\partial P(k)/\partial \delta_\br$ and $\partial P(k)/\partial
\tau_{Wij}$ are the responses of the power spectrum to the super-survey
modes, $\delta_\br$ and $\tau_{Wij}$, respectively. Here we consider the
power spectrum responses at the leading order of the super-survey modes,
or in other words we ignored the responses at the higher orders of
$O(\delta_\br^2,\tau_W^2)$. The ensemble average of the above power
spectrum, which is equivalent to the average of the power spectrum
estimator for different survey regions, is
\begin{equation}
 \avrg{P(\bk;\delta_\br,\tau_W)}
  = P(k)+\frac{\partial P(k)}{\partial 
 \delta_\br}\avrg{\delta_\br} + \frac{\partial P(k)}{\partial
 \tau_{Wij}}\avrg{\tau_{Wij}}=P(\bk),
\end{equation}
where we used $\avrg{\delta_\br}=\avrg{\tau_{Wij}}=0$. Thus the ensemble
average of the power spectrum estimator recovers the true power spectrum
in the global universe.  

Now let us consider the covariance matrix of the power spectrum
estimator (Eq.~\ref{eq:ps2d_est}): 
%
\begin{equation}
  C(\bk,\bk')=\avrg{\hat{P}(\bk)\hat{P}(\bk')}- P(\bk) P(\bk').
\end{equation}
Inserting Eq.~(\ref{eq:ps_resp}) into the above equation leads us to
find a formal expression of the super-sample covariance due to 
$\delta_\br$ and $\tau_W$:
\begin{equation}
 C^{\rm SSC}(\bk,\bk')=\sigma_\br^2\frac{\partial
  P(k)}{\partial \delta_\br}
  \frac{\partial
  P(k')}{\partial \delta_\br}
  +\avrg{\tau_{Wij}\tau_{Wlm}}\frac{\partial P(k)}{\partial \tau_{Wij}}
  \frac{\partial P(k')}{\partial \tau_{Wlm}},
  \label{eq:cov_response}
\end{equation}
where we have assumed $\avrg{\delta_\br\tau_{Wij}}\simeq 0$ for a
reasonably symmetric survey window.

Following the formulation in \citet{TakadaHu:13} \citep{Lietal:14b}, we
advocate that the squeezed trispectrum can be characterized by the
responses of the power spectrum to the super-survey modes as
\begin{equation}
 \lim_{q\rightarrow 0}\left[
T^{\rm PT}(\bk,-\bk+\bq,\bk',-\bk-\bq)-T^{\rm PT}(\bk,-\bk,\bk',-\bk')
						 \right]\simeq  P^L(q)
 \left[\frac{\partial P(k)}{\partial \delta_\br}+\tau_{Wij}
 \frac{\partial P(k)}{\partial \tau_{Wij}}\right]
 \left[\frac{\partial P(k')}{\partial \delta_\br}+\tau_{Wlm}
 \frac{\partial P(k')}{\partial \tau_{Wlm}}\right],
\end{equation}
where the Fourier modes $\bq$ are super-survey modes satisfying $k,
k'\gg q$.
Using the perturbation theory \citep{Bernardeauetal:02}, we can compute
the squeezed trispectrum contribution: 
\begin{eqnarray}
 C^{\rm
  SSC}(\bk,\bk')&=&\frac{1}{V_W^2}\int\!\!\frac{\mathrm{d}^2\bq}{(2\pi)^3}|\tW(\bq)|^2
  \left[T^{\rm PT}(\bk,-\bk+\bq,\bk',-\bk'-\bq)-T^{\rm
   PT}(\bk,-\bk,\bk',-\bk')\right].
  \label{eq:exp_ssc}
\end{eqnarray}
$T^{\rm PT}$ is the tree-level trispectrum, defined as
\begin{equation}
 \avrg{\delta(\bk_1)\delta(\bk_2)\delta(\bk_3)\delta(\bk_4)}_c
  =(2\pi)^3\delta_D^3(\bk_{1234})T^{\rm PT}(\bk_1,\bk_2,\bk_3,\bk_4).
\end{equation}
where
\begin{eqnarray}
   T^{\rm PT}(\bk_1,\bk_2,\bk_3,\bk_4)&=&
4\left[F_2(\bk_{13},-\bk_1)F_2(\bk_{13},\bk_2)
P^{\lin}(k_{13})P^{\lin}(k_1)P^{\lin}(k_2)+11 {\rm perm.}
\right]\nonumber\\
&&+6\left[F_3(\bk_1,\bk_2,\bk_3)
P^{\lin}(k_1)P^{\lin}(k_2)P^{\lin}(k_3)+\mbox{3 perm.}
	\right],
\label{eq:trisp_pt}
\end{eqnarray}
with the Fourier kernels defined as
\begin{eqnarray}
 F_2(\bk_1,\bk_2)&=&\frac{5}{7}+\frac{1}{2}\left(
					    \frac{1}{k_1^2}+\frac{1}{k_2^2}\right)
 (\bk_1\cdot\bk_2)+\frac{2}{7}\frac{(\bk_1\cdot\bk_2)^2}{k_1^2k_2^2},
\label{eq:f2}
\end{eqnarray}
and the definition of $F_3$ is given by Eq.~(31) in \citet{TakadaHu:13},
but the term involving $F_3$ is not relevant for the following
calculation. 

Inserting Eqs.~(\ref{eq:trisp_pt}) and (\ref{eq:f2}) into
Eq.~(\ref{eq:exp_ssc}) leads to
 \begin{eqnarray}
 C^{\rm SSC}(\bk,\bk')
 &\simeq & \frac{1}{V_W^2}
  \int\!\!\frac{\mathrm{d}^3\bq}{(2\pi)^3}|\tW(\bq)|^2
4 P^{L}(q)\left[
P^{L}(k) F_2(\bq,-\bk) 
+ 
P^{L}(|\bk-\bq|) F_2(\bq,\bk-\bq) \right]
\nonumber\\
 && \hspace{10em}\times
  \left[  P^{L}(k') F_2(\bq,\bk') +
 P^{L}(|\bk'+\bq|) F_2(-\bq,\bk'+\bq) \right],\nonumber\\
\end{eqnarray}
To further proceed the calculation, we need to care the fact that the
mode coupling kernel $F_2$ has a pole. More especially, under the fact
$k, k'\gg q$, we need to use the following expansion such as
\begin{eqnarray}
 P^L(|\bk-\bq|) F_2(\bq,\bk-\bq) \simeq
  \left[P(k)-\frac{\partial P(k)}{\partial k}(\bk\cdot\bq)\right]
  \left[\frac{5}{7}+\frac{1}{2}\left(\frac{1}{q^2}+\frac{1}{k^2}\right)
   \left(\bk\cdot\bq-q^2\right)+\frac{(\bk\cdot\bq-q^2)^2}{q^2k^2}
  \right].
\end{eqnarray}
Then we can find that the super-sample covariance can be computed as
\begin{eqnarray}
  C^{\rm SSC}(\bk,\bk')&\simeq&
  \sigma_\br^2\left[\frac{47}{21}-\frac{1}{3}\frac{\partial \ln P(k)}{\partial \ln k}\right]
  \left[\frac{47}{21}-\frac{1}{3}\frac{\partial \ln P(k')}{\partial \ln
   k'}\right]P^L(k)P^L(k')\nonumber\\
  && + \avrg{\tau_{Wij}\tau_{Wlm}}\hk_i\hk_j\hk_l^\prime\hk_m^\prime
  \left[\frac{8}{7}-\frac{\partial \ln P(k)}{\partial \ln k}\right]
  \left[\frac{8}{7}-\frac{\partial \ln P(k')}{\partial \ln k'}\right]P^L(k)P^L(k'),
\label{eq:cov_ssc_pt}
\end{eqnarray}
where $\hat{\bk}=\bk/k$.
To arrive at this equation, we used the following identities for the
$\bq$-integration:
\begin{eqnarray}
 \int\!\!\frac{\mathrm{d}^3\bq}{(2\pi)^3}~|\tW(\bq)|^2P^L(q)&=&\sigma_\br^2 \\
 \int\!\!\frac{\mathrm{d}^3\bq}{(2\pi)^3}~|\tW(\bq)|^2P^L(q)q_iq_j&=&
  \int\!\!\frac{\mathrm{d}^3\bq}{(2\pi)^3}~|\tW(\bq)|^2P^L(q)\left[
\left(q_iq_j-\frac{\delta^K_{ij}}{3}\right)+\frac{\delta^K_{ij}}{3}
									   \right]
  \nonumber\\
 &=&\avrg{\delta_\br\tau_{Wij}}+\frac{\delta^K_{ij}}{3}\sigma_{\br}^2\nonumber\\
 &\simeq&   \frac{\delta^K_{ij}}{3}\sigma_{\br}^2\\
 \int\!\!\frac{\mathrm{d}^3\bq}{(2\pi)^3}~|\tW(\bq)|^2P^L(q)q_iq_jq_lq_m&\simeq
 &
  \int\!\!\frac{\mathrm{d}^3\bq}{(2\pi)^3}~|\tW(\bq)|^2P^L(q)\left[
     \left(q_iq_j-\frac{\delta^K_{ij}}{3}\right)
     \left(q_lq_m-\frac{\delta^K_{lm}}{3}\right)
+\frac{\delta^K_{ij}\delta^K_{lm}}{9}
	   \right]
  \nonumber\\
 &=&\avrg{\tau_{Wij}\tau_{Wlm}}+\frac{\delta^K_{ij}\delta^K_{lm}}{9}\sigma_{\br}^2.
\end{eqnarray}
%
Note that we also used the fact that terms involving the moments with an
odd power of $q_i$ or equivalently an odd power of $k_i$ are vanishing
under the parity invariance conditions of $\bk\leftrightarrow -\bk$ and
$\bk'\leftrightarrow -\bk'$.

Comparing Eqs.~(\ref{eq:cov_response}) and (\ref{eq:cov_ssc_pt}) leads
us to find that the power spectrum response can be given as 
\begin{eqnarray}
 P(\bk;\delta_\br,\tau_W)\simeq P(k)
  +\delta_\br\left[\frac{47}{21}-\frac{1}{3}\frac{\partial \ln P(k)}{\partial \ln
    k}\right]P(k)
  +\hk_i\hk_j\tau_{Wij}\left[\frac{8}{7}-\frac{\partial \ln P(k)}{\partial \ln k}\right]P(k).
\end{eqnarray}

\bibliography{sst}

\end{document}